\documentclass[12pt]{article}

\usepackage{amsfonts, amsmath}

\newcommand{\be}{\begin{equation}}
\newcommand{\ee}{\end{equation}}

\begin{document}
\title{\bf Some  Comments on Dynamical Character of Cosmological Constant  and
GUP} \thispagestyle{empty}

\author{Alexander E. Shalyt-Margolin\hspace{1.5mm}\thanks
{e-mail: a.shalyt@mail.ru, alexm@hep.by}}
\date{} \maketitle
 \vspace{-25pt}
{\footnotesize\noindent  National Centre of Particles and High
Energy Physics, Bogdanovich Str. 153, Minsk 220040, Belarus\\
{\ttfamily{\footnotesize
}}

\rm\normalsize \vspace{0.5cm}
\begin{abstract}
In this paper the results obtained by Minic and his colleagues on the
uncertainty relation of the pair "cosmological constant - volume of
space-time", where cosmological constant is a dynamical quantity,
are reconsidered and generalized proceeding from a more  natural viewpoint.
It is demonstrated that on the basis of simple and natural assumptions
this relation may be understood with the help of the uncertainty relation for the
pair "energy - time". Since the latter is generalized at Planck's scales
(Early Universe)- GUP, the first one may be generalized in a similar
way. This means that we can suggest GUP for the pair "cosmological constant -
space-time volume". Here the relation is derived in the explicit form, and also
some  implications are considered.
\end{abstract}
\newpage
The Cosmological Constant Problem is basic for modern fundamental physics.
There are three principal questions associated with this problem.
\\ (a) Why the cosmological constant is nonzero?
\\(b) Why this constant is so small, being lower than the expected
theoretical value by a factor of $\sim10^{123}$?
\\(c )And why its actual value conforms well to the critical
density of vacuum energy $\rho_{c}$?
\\ Besides, the Cosmological Constant (Vacuum Energy Density) Problem
is closely connected with the Dark Energy Problem  that  has become
one of the key physical problems in basic research. Numerous works
and review papers on this problem have been published in the last
10-15 years \cite{Dar1}. And a great number of approaches
to this problem have been proposed: scalar field models
(quintessence model, K-essence, tachyon field, phantom field,
dilatonic, Chaplygin gas) \cite{Quint1},
\cite{K1},\cite{Tach1},\cite{Phant1},\cite{Dil1}, \cite{Chapl1},
braneworld models \cite{Brane1},dynamic approaches to the
cosmological constant $\Lambda$ \cite{Dyn1}, anthropic selection
of $\Lambda$ \cite{Antr1}, etc.
\\ At the same time, it should be noted that Cosmological Constant (Vacuum Energy)
persists to be the main candidate to play a role of Dark Energy. But we still
have no intelligible answers for the above questions. Because of this, any
progress in this direction is of particular value. By author's opinion,
most interesting in this respect are the works \cite{Min1}--\cite{Min4}.
Specifically, of great interest is the Uncertainty Principle derived in these works
for the pair of conjugate variables $(\Lambda,V)$:
\begin{equation}\label{CC1}
\Delta\Lambda\, \Delta V \sim \hbar,
\end{equation}
where $\Lambda$ is the vacuum energy density (cosmological
constant). It is a dynamical value fluctuating around zero;
$V$ is the space-time volume. Here the volume of space-time $V$
results from the Einstein-Hilbert action \cite{Min2}:
\begin{equation}\label{CC2}
S_{EH}\supset \Lambda \int d^{4}x \sqrt{-g}=\Lambda V
\end{equation}
In this case "the notion of conjugation is well-defined, but
approximate, as implied by the expansion about the static
Fubini--Study metric" (Section 6.1 of \cite{Min1}). Unfortunately,
in the proof per se (\ref{CC1}), relying on the procedure with a
non-linear and non-local Wheeler--de-Witt-like equation of the
background independent Matrix theory, some unconvincing arguments
are used, making it insufficiently rigorous (Appendix 3 of
\cite{Min1}). But, without doubt, this proof has a significant
result, though failing to clear up the situation.
\\Let us attempt to explain (\ref{CC1})(certainly at an heuristic level)
using simpler and more natural terms involved  with the other, more well-known,
conjugate pair $(E,t)$ - "energy - time". We use the designations of
\cite{Min1},\cite{Min2}. In this way a four-dimensional volume will be
denoted, as previously, by $V$.
\\ Just from the start, the Generalized Uncertainty
Principle (GUP) is used. Then a change over to the Heisenberg Uncertainty Principle
at low energies will be only natural. As is known, the Uncertainty
Principle of Heisenberg at Planck's scales (energies) may be extended to the
Generalized Uncertainty Principle. To illustrate, for the conjugate pair
"momentum-coordinate" $(p,x)$ this has been noted in many works
\cite{Ven1}--\cite{Magg1}:
\begin{equation}\label{CC3}
\triangle x\geq\frac{\hbar}{\triangle p}+\alpha^{\prime}
l_{p}^2\frac{\triangle p}{\hbar}.
\end{equation}
In \cite{Shalyt3},\cite{Shalyt9} it is demonstrated that
the corresponding Generalized Uncertainty Relation for the pair "energy
- time" may be easily obtained from
\begin{equation}\label{CC4}
\Delta t\geq\frac{\hbar}{\Delta E}+\alpha^{\prime}
t_{p}^2\,\frac{\Delta E}{ \hbar},
\end{equation}
where $l_{p}$ and $t_{p}$ - Planck length and time, respectively.
\\Now we assume that in the space-time volume $\int d^{4}x \sqrt{-g}=
V$ the temporal and spatial parts may be separated (factored out)
in the explicit form:
\begin{equation}\label{CC5}
V(t)\approx t \bar{V}(t),
\end{equation}
where $\bar{V}(t)$ - spatial part $V$.  For the expanding (inflation)
Universe such an assumption is quite natural.
Then it is obvious that
\begin{equation}\label{CC6}
\Delta V(t)=\Delta t \bar{V}(t)+ t \Delta\bar{V}(t)+\Delta t
\Delta \bar{V}(t).
\end{equation}
Now we recall that for the inflation Universe the scaling factor
is $a(t)\sim e^{Ht}$. Consequently, $\Delta\bar{V}(t)\sim \Delta t^{3}f(H)$,
where $f(H)$ is a particular function of Hubble's constant.  From (\ref{CC4})
it follows that
\begin{equation}\label{CC7}
\Delta t\geq t_{min}\sim t_{p}.
\end{equation}
However, it is suggested that, even though $\Delta t$ is satisfying
(\ref{CC7}), its value is sufficiently small in order that $\Delta V$
be contributed significantly by the terms containing $\Delta t$ to the power
higher than the first. In this case the main contribution on the right-hand side
of (\ref{CC6}) is made by the first term $\Delta t \bar{V}(t)$ only.
Then, multiplying the left- and right-hand sides of (\ref{CC4}) by $\bar{V}$ ,
we have
\begin{equation}\label{CC8}
\Delta V\geq\frac{\hbar \bar{V}}{\Delta E}+\alpha^{\prime}
t_{p}^2\,\frac{\Delta E \bar{V}}{ \hbar}= \frac{\hbar}{\Delta
\Lambda}+\alpha^{\prime} t_{p}^2 \bar{V}^{2}\frac{\Delta \Lambda}{
\hbar}.
\end{equation}
It is not surprising that a solution of the quadratic inequality
(\ref{CC8}) leads to a minimal volume of the space-time
$V_{min}\sim V_{p}=l_{p}^{3}t_{p}$ since (\ref{CC3}) and
(\ref{CC4})  result in minimal length $l_{min}\sim l_{p}$ and
minimal time $t_{min}\sim t_{p}$, respectively.
\\(\ref{CC8}) is of interest from the viewpoint of two limits:
\\1)IR - limit: $t\rightarrow \infty$
\\2)UV - limit: $t\rightarrow t_{min}$.
\\In the case of IR-limit we have large volumes $\bar{V}$  and  $V$ at
low $\Delta\Lambda$. Because of this, the main contribution in the right-hand side
of (\ref{CC8}) is made by the first term as great $\bar{V}$ in the second term
is damped by small $t_{p}$ and $\Delta \Lambda$.
Thus, we derive at
\begin{equation}\label{CC9}
\lim\limits_{t\rightarrow \infty}\Delta
V\approx\frac{\hbar}{\Delta \Lambda}
\end{equation}
in accordance with (\ref{CC1}) \cite{Min1}. Here, similar to \cite{Min1} ,
$\Lambda$ is a dynamical value fluctuating around zero.
\\And for the case (2) $\Delta\Lambda$ becomes significant
\begin{equation}\label{CC10}
\lim\limits_{t\rightarrow t_{min}}\bar{V}=\bar{V}_{min}\sim
\bar{V}_{p}=l_{p}^{3};  \lim\limits_{t\rightarrow
t_{min}}V=V_{min}\sim V_{p}=l_{p}^{3}t_{p}.
\end{equation}
As a result, we have
\begin{equation}\label{CC11}
\lim\limits_{t\rightarrow t_{min}}\Delta V = \frac{\hbar}{\Delta
\Lambda}+\alpha_{\Lambda} V_{p}^{2}\frac{\Delta \Lambda}{\hbar},
\end{equation}
where the parameter $\alpha_{\Lambda}$  absorbs all the
above-mentioned proportionality coefficients.
\\ For(\ref{CC11}) $\Delta \Lambda \sim
\Lambda_{p}\equiv\hbar/V_{p}=E_{p}/\bar{V}_{p}$.
\\ It is easily seen that in this case $\Lambda \sim M_{p}^{4}$, in
agreement with the value obtained using a naive  (i.e. without
super-symmetry and the like) quantum field theory
\cite{Wein1},\cite{Zel1}. Despite the fact that $\Lambda $ at
Planck's scales (referred to as $\Lambda(UV) $) (\ref{CC11}) is
also a dynamical quantity, it is not directly related to
well-known $\Lambda $ (\ref{CC1}),(\ref{CC9}) (called $\Lambda(IR)
$) because the latter, as opposed to the first one, is derived
from Einstein's equations
\begin{equation}\label{CC12}
R_{\mu \nu} - \frac{1}{2} g_{\mu \nu} R = 8\pi G_N \left( -
\Lambda g_{\mu \nu} + T_{\mu \nu} \right).
\end{equation}
However, Einstein's equations (\ref{CC12}) are not valid at the
Planck scales and hence $\Lambda(UV) $  may be considered as some
high-energy generalization of the conventional cosmological
constant, leading to $\Lambda(IR) $ in the low-energy limit.
\\In conclusion, it should be noted that the right-hand side of
(\ref{CC3}),(\ref{CC4}) in fact is a series.
Of course, a similar statement is true for (\ref{CC11}) as well.
 \\Then, we obtain a
system of the Generalized Uncertainty Relations for the Early Universe
(Planck scales) in the symmetric form as follows:
\begin{equation}\label{CC13}
\left\{
\begin{array}{lll}
\Delta x & \geq & \frac{\displaystyle\hbar}{\displaystyle\Delta
p}+ \alpha^{\prime} \left(\frac{\displaystyle\Delta
p}{\displaystyle p_{pl}}\right)\,
\frac{\displaystyle\hbar}{\displaystyle p_{pl}}+... \\ &  &  \\
\Delta t & \geq & \frac{\displaystyle\hbar}{\displaystyle\Delta
E}+\alpha^{\prime} \left(\frac{\displaystyle\Delta
E}{\displaystyle E_{p}}\right)\,
\frac{\displaystyle\hbar}{\displaystyle E_{p}}+...\\
  &  &  \\
  \Delta V & \geq &
  \frac{\displaystyle \hbar}{\displaystyle\Delta \Lambda}+\alpha_{\Lambda}
  \left(\frac{\displaystyle\Delta \Lambda}{\displaystyle \Lambda_{p}}\right)\,
  \frac{\displaystyle \hbar}{\displaystyle \Lambda_{p}}+...
\end{array} \right.
\end{equation}
The last of the relations (\ref{CC13}) may be important when
finding the general form for $\Lambda(UV) $, low-energy limit
$\Lambda(IR) $, and also may be a step in the process of
constructing future quantum-gravity equations, the low-energy
limit of which is represented by Einstein's equations
(\ref{CC12}).
\newpage
%References

\end{document}